\documentclass[11pt]{article}

\usepackage[table]{xcolor}
\usepackage{helvet}  
\usepackage[utf8]{inputenc}
\definecolor{darkgreen}{HTML}{148c82}
\definecolor{lightgreen}{HTML}{dffef3}  
\definecolor{magenta}{HTML}{ae1d82}
\usepackage{array}
\usepackage{colortbl}  

\usepackage{fancyhdr}

\usepackage{graphicx}%
\usepackage{textcomp}%
\usepackage{manyfoot}%

\usepackage{crimson}
\usepackage{xcolor}
\usepackage[switch]{lineno}
\usepackage{authblk}
\usepackage{booktabs}
\usepackage{tabularx}
\usepackage{rotating}

\usepackage{url}
\usepackage{tcolorbox}

\title{The Science Fiction Science Method}

\author[1,*]{Iyad Rahwan}
\author[2,*]{Azim Shariff}
\author[3,*]{Jean-Fran\c{c}ois Bonnefon}

\affil[1]{\footnotesize Center for Humans \& Machines, Max Planck Institute for Human Development, Berlin, Germany}
\affil[2]{Department of Psychology, University of British Columbia, Vancouver, Canada}
\affil[3]{Toulouse School of Economics, CNRS (TSM-R), University of Toulouse-Capitole, France}
\affil[*]{Correspondence: \texttt{rahwan@mpib-berlin.mpg.de}, \texttt{shariff@psych.ubc.ca}, 
\texttt{jean-francois.bonnefon@tse-fr.eu}
}

\begin{document}

\fancypagestyle{titlepagestyle}{
  \fancyhf{} 
  \fancyhead[C]{The definitive version of this article is published in \textit{Nature} (2025). It can be found at:
  \url{https://www.nature.com/articles/s41586-025-09194-6}
} 
  \renewcommand{\headrulewidth}{0.4pt} 
}

\date{}

\maketitle

\thispagestyle{titlepagestyle} 

\section*{Abstract}

Predicting the social and behavioral impact of future technologies, before they are achieved, would allow us to guide their development and regulation before these impacts get entrenched. Traditionally, this prediction has relied on qualitative, narrative methods. Here we describe a method which uses experimental methods to simulate future technologies, and collect quantitative measures of the attitudes and behaviors of participants assigned to controlled variations of the future. We call this method `science fiction science’. We suggest that the reason why this method has not been fully embraced yet, despite its potential benefits, is that experimental scientists may be reluctant to engage in work facing such serious validity threats as science fiction science. To address these threats, we consider possible constraints on the kind of technology that science fiction science may study, as well as the unconventional, immersive methods that science fiction science may require. We seek to provide perspective on the reasons why this method has been marginalized for so long, what benefits it would bring if it could be built on strong yet unusual methods, and how we can normalize these methods to help the diverse community of science fiction scientists to engage in a virtuous cycle of validity improvement.

\newpage

\section*{Main}

Imagine that behavioral scientists managed to predict the effect of social media on mental health and democratic life, before social media actually existed.
That is, imagine they designed simulations of what they thought social media might look like, and recorded participants' interactions with this speculative technology. 
They might have noted a tendency for participants to focus on upward social comparisons, leading to self-esteem issues \cite{braghieri2022social}, or a focus on moral outrage, resulting in exaggerated feelings of polarization \cite{van2024social}. 
These speculative findings could have guided us in designing and regulating social media with the benefit of foresight, instead of constantly playing catch-up with their  effects \cite{bail22socialmedia}.

Social media is just one example of a technology that could have benefited from early, speculative behavioral research before widespread deployment.
For example, when genetically modified foods were taken to market in the 1990s, they faced a strong public opposition that caught both the industry and policymakers off guard.
Companies and regulatory bodies had failed to anticipate the psychological concerns of the public about this technology \cite{frewer2004societal}, which led to a persistent cycle of mistrust that has not been fully resolved even 30 years later \cite{siegrist2020consumer}.
Today, many people  still manifest absolute moral opposition to genetically modified foods \cite{scott2016evidence}, although they understand very little about them \cite{fernbach2019extreme}. It is of course impossible to claim that history would have been different if behavioral scientists had conducted experiments on the acceptability of genetically modified foods before the technology was fully developed. 
However, the experience with genetically modified foods provided a valuable lesson, leading to the normalization of prospective social acceptability studies in other fields.
For example, people were polled about their reactions to some potential applications of nanotechnologies well before these applications became possible \cite{cobb2004public,lee2005public}.

In this perspective, we first observe that despite a clear and consensual need to predict the social and behavioral impact of future technologies, it remains uncommon to do so with the experimental methods and quantitative measures typical of contemporary behavioral science.
Consequently, we advocate for what we term \textit{science fiction science:} the application of the scientific method to the science fiction project of anticipating behavioral and social changes driven by a speculative technology.
After providing a more detailed definition of science-fiction science, we review past and current examples, illustrating its primary challenges: prospective validity and cost-benefit analysis. 
We discuss the principles and methods that have and could be used to overcome these challenges.
Our goals are to synthesize the reasons why we need science fiction science, and the reasons why it is so disorganized still; to foster a community of researchers under this unified banner; to legitimize the unconventional methods required for the unconventional enterprise of conducting behavioral experiments centered on technologies that do not yet exist; and to help the diverse community of science fiction scientists to engage in a virtuous cycle of methodological improvement and community growth.


\section*{Technology-driven futures}

New technologies transform societies by changing what people can do, what they actually do, and what they think is acceptable to do \cite{danaher2023mechanisms}.
For example, the introduction of a reliable birth control pill gave women unprecedented control over their reproductive lives, education and careers, transforming social norms around premarital sex, family planning, and women's participation in the workforce \cite{goldin2002power,bailey2006more}. 
In parallel, the technology that made organ donation easier by maintaining heart and lung function in brain-dead patients changed the way doctors and patient think of the ethics of ending life \cite{baker2013before}. 
Yet other technologies transformed society because they were \textit{not} accompanied by significant changes in behavior and social norms. 
For example, sophisticated tracking technologies allowed tech giants to accumulate vast amount of personal data, leading to a commodification of human experience, because people failed to adjust their behaviors and norms in a way that would have sustained privacy  \cite{shariff2021privacy, zuboff2019age}.

In all these examples, we have the benefit of hindsight about the impact of each technology. However, our capacity to now alter these impacts--if we wanted to--has become limited by behaviors, norms, and institutions now deeply entrenched. 
In contrast, we have much power to steer and regulate technologies which are in their early or speculative stage---but when a technology is at this speculative stage, we know little about the impact it may have on society, or how our actions may change this impact (see box \emph{Examples of policy debates about nascent or speculative technologies}).
This tension is known as the Collingridge control dilemma \cite{collingridge1980social}: the moment when we have the best chance to shape the social impact of a technology is also the moment when we know the least about what it will actually do and how.
One way out of this dilemma is to keep regulation as agile and flexible as possible, so that it can quickly adapt to our emerging understanding of a technology. 
Another escape from the dilemma is to attempt to predict, as early as possible, the different futures that may unfold under the influence of a given technology:
For example, what futures are the most likely if the technology is left unregulated, what regulations would be the most acceptable to the people of the future, and what futures may further unfold as a result of these regulations.

\noindent
\begin{tcolorbox}[colback=gray!20!white, colframe=black, float, floatplacement = t, title=Examples of policy debates about nascent or speculative technologies.]
\scriptsize
\begin{description}
    \item[Connected and autonomous vehicles (CAVs)] Autonomous driving technology is still being tested and fully autonomous cars are confined to specific cities. Yet, a debate ensued early on about the ethical implications of intelligent machines making life and death decisions in unavoidable accident scenarios \cite{goodall2014ethical}. These debates have led to significant mobilization by Civil Society \cite{hess2020incumbent}, and specific policy recommendations \cite{bonnefon2020ethics}, some of which have informed legislation \cite{kriebitz2022german}.
    \item[Social credit systems] Governments do not yet have the capacity to deploy AI systems that would monitor every behavior of all their citizens in real time, in order to calculate and release social scores; but the behavioral, social and political impact of these hypothetical systems is already the object of much speculation \cite{mac2019chinese,orgad2019dystopia,tirole2021digital,purcell2023humans}, to the point where the European Union is considering a preemptive ban of this technology for its member governments \cite{EC2021AIAct}.
    \item[Embryo screening] Fertility clinics are only in the very early stages of providing parents with the option to choose or reject specific genetic traits in their offspring \cite{turley2021problems}. However, bioethicists are already considering the ethical trade-offs, social dilemmas, and collective action problems that may arise if parents gain the ability to selectively determine cognitive, moral, physical, and immune traits \cite{anomaly2020creating, anomaly2020great, gyngell2015stocking}.
    \item[Ectogenesis] Medical research has not yet achieved ectogenesis (i.e., the artificial gestation of human foetuses), but the field of speculative bioethics is already debating whether this potential technology might end gender-based oppression, or exacerbate other inequalities, and to which extent it could disrupt social and legal norms around motherhood and employment rights \cite{Cavaliere2020-CAVEAG,Hooton2022-AWT,Horn2023-EIC,MacKay2020-TYR}.
\end{description}
\end{tcolorbox}

Anticipating technology-driven futures has long been the project of \textit{futures studies,} which primarily use qualitative methods such as Delphi surveys and other techniques aimed at organizing communication between expert panelists \cite{fergnani2019mapping}. 
Future studies typically attempt to compose various scenarios describing the future, based on expert intuition, and then work backwards to imagine which actions or events may lead to these scenarios \cite{kuosa2011evolution}.
Besides futurists, ethicists and legal scholars also show interest in the social and moral impact of speculative technologies, and they also primarily use qualitative methods, such as normative analysis and case studies \cite{danaher2022technology,hopster2023technology}. 
What is much less common is for scientists to weigh in on these debates by producing quantitative data extracted from experimental methods and behavioral measures \cite{brey2017ethics}. 
These methods are routinely used to examine the impact of existing technologies, but their presence sharply diminishes when turning to future technologies.
In other words, there seems to be no well-established field that would be the quantitative, experimental, behavioral counterpart to futures studies and other qualitative explorations of the impact of speculative technologies.

\section*{Science fiction science}

Let us summarize our problem so far.
We start from a technology that is not available yet, but is likely to become available in the future, with some ill-defined capabilities and limitations. 
We believe it may transform individual behaviors, social norms, and institutions, with a mixture of desirable and undesirable effects. 
The potential magnitude of these effects is  large enough for us to worry about them even before the technology is achieved, in the hope that we can steer its development toward the most desirable effects, and prepare to regulate against its remaining undesirable effects.
The challenge is that  because the technology does not exist yet, we have no behavioral data about what people will think and do when it becomes available, no controlled experiments, no tentative quantification of its positive and negative effects.

To collect such data, we need to develop methods and principles for a speculative behavioral science.
We propose to call this method \textit{science fiction science} (or sci-fi-sci for short), because, at its core, it consists of applying the scientific method to the science-fiction project.
Indeed, science-fiction writers provide elaborate thought experiments about the social and behavioral impact of future technologies. 
As science fiction author Frederik Pohl once wrote, paraphrasing Isaac Asimov: `Somebody once said that a good science-fiction story should be able to predict not the automobile but the traffic jam. \cite{pohl68asimov}'
Sci-fi-sci is an attempt to turn the thought experiments of science fiction into behavioral experiments: science fiction tells the stories of future humans, but sci-fi-sci attempts to study them in the lab.

More precisely, sci-fi-sci is the application of the scientific method to anticipate and study the behavioral, psychological, attitudinal, and social impacts of speculative technologies that do not yet exist. 
By recruiting present-day participants and immersing them into controlled experimental simulations of possible futures, sci-fi-sci obtains quantitative measures of their thoughts, attitudes, and behaviors. 
This approach employs rigorous experimental methods—including control and treatment conditions with independent and orthogonal variable manipulation, to not only survey attitudes but also measure direct behaviors as participants interact with simulated future technologies. 
By doing so, sci-fi-sci aims to uncover potential sociotechnical benefits and harms, and to offer insights into how different factors (including marketing) can influence public perception and behavior. 
It serves as the quantitative, experimental, and behavioral counterpart to Futures Studies and other qualitative explorations of the impact of speculative technologies.

Our own research provides a notable example of sci-fi-sci (see \textit{Case Study A} box for further details). 
Between 2016 and 2018, we published several articles on fully autonomous vehicles, a technology that was widely anticipated, but had not yet been achieved. In these articles, we surveyed participants on the ethical preferences they would want these vehicles to follow in the event of an unavoidable collision, where the vehicle must decide which road users to spare and which to sacrifice  \cite{bonnefon2016social,shariff2017psychological, awad2018moral}. 
In particular, we used a conjoint design which allowed for detecting the trade-offs participants were willing to make for these techno-ethical choices, a method which may be particularly useful to measure attitudes and behaviors driven by speculative technologies.

\noindent
\begin{tcolorbox}[colback=gray!20!white, 
colframe=black, title=Case Study A: Ethical dilemmas of autonomous vehicles (AVs)]
\scriptsize
\begin{description}
    \item[Period of Sci-Fi-Sci studies:] 2016--today
    \item[Sci-fi-sci Question:] How do citizens and consumers wish AVs to prioritize the safety of different road users in unavoidable accidents? Technology ethicists \cite{lin2016ethics} and transportation experts \cite{goodall2014ethical} were already discussing such possibilities before the mid-2010s.

    \item[Technological Plausibility:] Many proof-of-concept tests took place through the DARPA Urban Challenge in 2007 (Technology Readiness Level=6). By 2015, various US states even allowed AV testing on public roads (TRL=7--8). 

    \item[Temporal Proximity:] Spurred by early success, the market invested billions of dollars into AV technology development, suggesting that the technology was very proximal--even while estimates varied. Indeed, by 2024, Waymo had deployed its fully autonomous taxi service in the City of San Francisco (TRL=9--10).

    \item[Magnitude of Effect:] AVs could have substantial socioeconomic effects such as changing urban land use and altering commuter behavior or even the value of time \cite{milakis2019long}. Many of these outcomes, often based on computational simulation, are highly sensitive to modeling assumptions \cite{soteropoulos2019impacts}, and are thus difficult to study behaviorally. In contrast, it is more feasible to study consumer attitudes that may shape the early adoption of AVs, and how consumers and citizens may react to different design features and regulatory regimes.
    
    \item[Sample of sci-fi-sci studies:] Starting in 2016, a series of studies attempted to anticipate how people would react to AV accident dilemmas. Initial studies used text vignettes \cite{bonnefon2016social} highlighting the disconnect between the preferences of citizens (AVs should save as many lives as possible) and those of consumers (AVs should prioritize passengres). The Moral Machine experiment \cite{awad2018moral} used a conjoint design with visualizations of binary choices, crowdsourcing over 40 million decisions from people world wide, and highlighting cross cultural differences. More recent studies used Virtual Reality to situate participants in the expected reality of a driverless car carrying out complex moral decisions \cite{benvegnu2021virtual,sutfeld2019does,faulhaber2019human}.

    \item[Subsequent relevance:] By engaging millions of citizens world-wide, the Moral Machine experiment itself has contributed to a significant Civil Society debate around AV ethics, which in turn informs policy making \cite{hess2020incumbent}. More broadly, sci-fi-sci studies in this context have informed specific policy recommendations \cite{bonnefon2020ethics,EU:Trustworthy}, and subsequent legislations \cite{kriebitz2022german}.
\end{description}
\label{fig:cast-study-A}
\end{tcolorbox}

In hindsight, this was sci-fi-sci research, in the sense that we asked our participants to imagine a future technology, assigned them to various experimental treatments corresponding to hypothetical regulations of that technology, and recorded behavioral measures such as their support for the government that enacted the regulation, or their intention to purchase the regulated technology. 
These articles and many other subsequent behavioral articles about autonomous vehicles allowed policy debates to take place ahead of technology developments: 
Fully autonomous vehicles are still not available for purchase, but thanks to sci-fi-sci, we have made much progress on their ethical regulation \cite{bonnefon2020ethics, luetge2017german, santoni2021european}, and we are less likely to be blindsided as a society by their behavioral implications \cite{adnan2024exploring}. 
In like vein, research on service robots has a long history of attempting to predict the behavioral interactions people may have with futuristic robots, as well as the social implications of their introduction \cite{tussyadiah2020review,zeng2024retrospective} (see \textit{Case Study B} box for further details).

Before we consider other sci-fi-sci examples, we should first comment on the challenge of conducting a systematic literature review, or assessing the impact of this approach on technology development and regulation. 
Sci-fi-sci, unlike futures studies (a well-established field with dedicated journals, recognized methods, and searchable keywords) lacks easy retrievability. 
It is not easily searchable, nor centralized in specific journals or conferences.
Moreover, evaluating the influence of sci-fi-sci on technological development and regulation can be problematic, because tech companies (and sometimes, policymakers) lack transparency about how they incorporate behavioral research in their decision making.
This problem is compounded by the fact that experimental research on the impact of speculative technologies may be increasingly shaped or conducted by tech companies with few incentives to make their findings public \cite{benkler2019don}.
Given the applied importance of predicting the behavioral and social effects of emerging technologies, especially at a time when advances in Artificial Intelligence spawn both hope and fear \cite{cave2019hopes}, we need to understand the relative reluctance of behavioral scientists to engage in the experimental exploration of technology-driven futures. In other words, why has sci-fi-sci not yet become a well-established field?

\noindent
\begin{tcolorbox}[colback=gray!20!white, colframe=black, 
title=Case Study B: Cooperating with autonomous agents]
\scriptsize
\begin{description}
    \item[Period of Sci-Fi-Sci studies:] 2000s--today
    \item[Sci-fi-sci Question:] How should an autonomous agent (e.g. robot, software agent) prioritize its own interests? This question was posited by sci-fi author Isaac Asimov in his 3rd law: `A robot must protect its own existence as long as such protection does not conflict with the First or Second Law' \cite{asimov1950robot}. In a 2011 article titled `What matters to a machine?', AI pioneer Drew McDermott imagined a robot  tempted to break an ethical rule to further its owners interests \cite{mcdermott2011matters}.

    \item[Technological Plausibility:] Early software agents and robots were mere tools performing specific tasks. They did not exhibit autonomy, nor did they face situations in which their goals conflicted with humans. By the early 2000s, we had various experimental demonstrations of autonomous social robots \cite{hirai1998development,ishiguro2001robovie,breazeal2004designing}, and software agents that manage business processes \cite{jennings2000autonomous} (TRL=5--7). Scenarios requiring such agents to negotiate their own interests with humans became increasingly plausible.

    \item[Temporal Proximity:] Approximately two decades later, software agents became mainstream \cite{reijers2021business}  (TRL=9--10), and corporate investment in humanoid robotics reached USD 2.43 billion in 2023 \cite{humanoid} (TRL=7--8). This led AI experts to demand that autonomous machines must learn to find common ground and cooperate with each other and with humans \cite{dafoe2021cooperative}.

    \item[Magnitude of Effect:] The long-term socioeconomic effects of autonomous agents are difficult to predict \cite{acemoglu2019automation}.  However, it is feasible to study the early dynamics of communication and cooperation between humans and autonomous agents when their interests are not fully aligned.
    
    \item[Sample of sci-fi-sci studies:] About a decade ago, a series of studies explored whether autonomous agents, tasked with maximizing their own interest, can establish stable cooperation with humans \cite{crandall2018cooperating}. These experiments adapted paradigms from behavioral economics, such as the Repeated Prisoner's Dilemma. These studies revealed that humans were less likely to cooperate with benevolent machine agents compared to humans \cite{ishowo2019behavioural}, because humans consider it acceptable to exploit them \cite{karpus2021algorithm}. These tendencies are influenced by visual features of autonomous agents \cite{oudah2024perception} as well as signals about emerging human-machine norms \cite{makovi2023trust}. Related work explored situations in which machine agents can exert authority over humans, e.g. as managers \cite{dong2024toward}, and identified moral hazards that may arise because people exhibit obedience to the authority of a robot boss \cite{aroyo2018will}.

    \item[Subsequent relevance:] The studies cited above may have seemed highly speculative just a few years ago. But this all changed in the last two years with the sudden and rapid rise of conversational agents like ChatGPT \cite{bubeck2023sparks}, and their numerous autonomous (`agentic') implementations \cite{wu2023autogen}. Suddenly, we live in a world where we interact frequently with autonomous customer service agents acting on behalf of other organizations \cite{bamberger2023generative}. The challenge of establishing cooperation between humans and autonomous agents is already pervasive, but fortunately we had a head start.
\end{description}
\label{fig:cast-study-B}
\end{tcolorbox}

\section*{Challenges for science fiction science}

Behavioral scientists who seek to inform policymaking and technological regulation need to tackle the challenge of the ecological validity of their experiments (i.e., the likelihood that the findings they obtain in the lab can predict behavior in the real world) as well as the temporal validity of their findings (i.e., the endurance of their ecological validity into a changing future) \cite{lazer2020computational,lazer2021meaningful, munger2019limited,munger2023temporal}. 
These challenges are especially problematic for sci-fi-sci experiments. 
Sci-fi-sci experiments cannot have ecological validity in a strict sense because there is no `real world' that the studies seek to generalize to. 
The world they try to generalize to does not exist yet, and may in fact never exist. 
Indeed, the challenge of temporal validity is inverted for sci-fi-sci experiments. 
For traditional experiments, the challenge of temporal validity is that ecological validity decays over time, from the present moment onward. 
For sci-fi-sci experiment, the hope is that their ecological validity will increase over time, in the sense that their findings will one day reflect real-world behavior, once their target technology is deployed for real.

Threats to such prospective ecological validity can occur in three areas. 
First, participants from the present may fail to simulate the behavior of actual users of the technology. 
We know, for example, that people can have a hard time predicting their future emotional states \cite{wilson2003affective}, even when they are trying to picture situations they know well. 
A sci-fi-sci experiment may pose an even greater challenge to participants, if it requires them to imagine their cognitive and emotional reactions to an unfamiliar future technology in an unfamiliar future context.
Second, the depiction of the technology used by experimenters might be significantly different from the actual version once it is developed.  
For example, 20 years ago, prospective studies of the acceptability of nanotechnologies asked people how comfortable they would be with the nano-augmentation of cognitive capacities \cite{cobb2004public,lee2005public}. 
Given that nanotechnologies have not developed in that direction, we can now recognize that the prospective validity of these questions was low.
Third, the social context of the experiment may differ significantly from the social context at the time the technology is developed.
Changes in social context can significantly impact people’s attitudes and behaviors towards technology. 
For example, the COVID-19 pandemic increased acceptance of care robots \cite{schonmann2024contagious}, and shifts in religiosity can affect attitudes towards assisted reproduction technologies \cite{inhorn2008assisted}. 
Unexpected changes in social context between the sci-fi-sci experiment and the actual deployment of the technology may thus threaten prospective validity.

These are formidable challenges. 
Without a set of accepted principles and methods to address them, behavioral scientists may have been dissuaded from producing what would otherwise have been useful research. 
In fact, it is plausible that the absence of established approaches has created a self-perpetuating cycle that prevents behavioral scientists from conducting experiments on the effects of future technologies. 
Concerns about methodological issues may have made researchers hesitant to predict these effects---and this hesitation, in turn, may have prevented the formation of a diverse community of behavioral scientists dedicated to developing the principles and methods needed to study future technologies.

\begin{figure*}
\includegraphics[width = \textwidth]{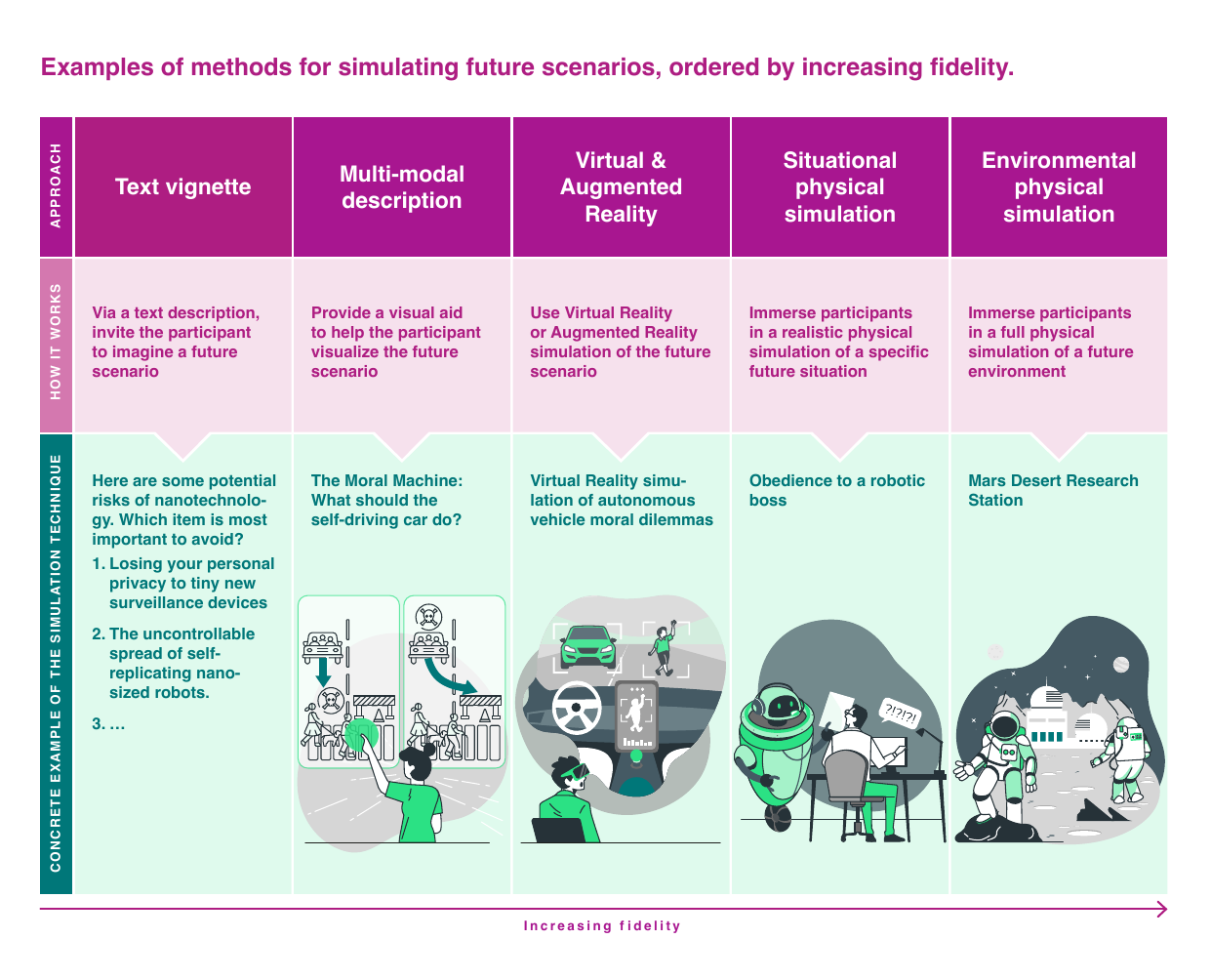}
\caption{\textit{Examples of methods for simulating future scenarios, ordered by increasing fidelity.}}\label{fig:fidelity}
\end{figure*}

Nevertheless, researchers from several fields have experimented with methods that may minimize threats to prospective validity (see Figure \ref{fig:fidelity}). 
For instance, many studies devised methods to simulate the experience of a technology, in order to bring participants psychologically closer to the future that the researchers are attempting to study. 
Often, this type of simulation has been limited to vignettes: The technology is described to participants, leaving to their imagination how it would feel to use it. 
This has sometimes been the only feasible option, such as in situations like neuro-enhancement drugs, where the technology cannot be simulated otherwise \cite{dinh2020public,mihailov2021pills,sattler2022neuroenhancements,lucas2024moral}. 
However, the more that contextual factors are left to the imagination of the participants, the more participants will fill in the gaps themselves, leading to unwanted variation. 
For example, some participants may project into the experiment the context of some science-fiction works they are familiar with \cite{laakasuo2018makes}. 
In order to limit such unwanted variations, experimenters need to provide as much contextual information they can. One way to move beyond vignettes and to improve the consistency and verisimilitude of the simulated experience is to use of mock versions of futuristic apps \cite{salganik2006experimental, longoni2022artificial,kobis2023artificial}.
Although these mockups cannot execute the full range of capabilities of the technology they are simulating, they embed an anticipated future technology within a context that participants are accustomed to. 
For example, Longoni and Cian \cite{longoni2022artificial} presented participants with a mobile app that purported to provide recommendations of an AI master chocolatier. 
Simulated apps may also allow to use immersive behavioral measures, allowing participants to `stay in character' as their future selves, rather than risking a jarring transition to standard survey questions that could undermine the effect of the simulation.

Researchers are also increasingly turning to virtual and augmented reality to further increase immersion. 
For example, the moral dilemmas of autonomous vehicles were initially studied through vignettes with static images \cite{bonnefon2016social}, then later with interactive computer graphics \cite{awad2018moral}, and most recently with virtual reality \cite{benvegnu2021virtual,sutfeld2019does,faulhaber2019human} in order to better situate participants within the expected reality of a world with pervasive driverless cars carrying out complex moral decisions. 
Other studies have used digital and even physical avatars to simulate the interaction with hypothetical AI-powered systems, advanced beyond the capabilities of current AI.
For example, researchers have used `Wizard of Oz' techniques, where human confederates impersonated AI chatbots with responsive abilities that were at the time impossible for the chatbots to generate themselves \cite{riek2012wizard}. 
This technique has also been used to give research participants the experience of interacting with speculative humanoid robots, for example robot bosses whose blinking and neck movements are autonomously generated, but whose decisions and responses are controlled in real-time by experimenters monitoring the interaction via cameras \cite{aroyo2018will}.
Finally, at the far end of the fidelity spectrum, some research made use of extremely elaborate staged settings, inaccessible to most researchers---for example, the Mars missions simulations \cite{bishop2010fmars, alfano2018long,riva2022social} involved purpose-built habitats to mimic the confinement, isolation, and resource-scarcity of a future micro-society in a new environment.

These studies showcase some methodological solutions that researchers have found to the problem of running experiments on the future, and it remains to be seen how successful these attempts can be. 
In addition, while the core of sci-fi-sci revolves around experimental studies involving human participants, we recognize the valuable role that computational methods such as Monte Carlo simulations and agent-based modeling can play in enriching our understanding of the future. 
These methods allow researchers to model complex systems and explore how assumptions about human behavior and technological adoption might play out over time. 
In the same spirit, AI simulations using bots programmed with different behavioral assumptions may sometimes serve as proxies for human participants, especially when exploring large-scale social dynamics that are impractical to study in laboratory settings. 
Integrating these computational approaches can complement the experimental data obtained from human participants, offering a richer view of potential futures shaped by speculative technologies.

As the sci-fi-sci literature grows, the community will trade perspectives and accumulate data on which methods work best for which purposes, and when. When is complex world-building necessary? When do simple vignettes suffice and when are laborious virtual or physical simulations beneficial? Conversely, when do these simulations risk creating demand effects linked to the experimenters' own vision of the future? When does prospective validity benefit from having immersive measures in addition to immersive stimuli?
But the solution space also has considerable room to grow. 
As noted, these methods are scattered across various disciplines, from psychology to experiential ethnographic futures \cite{candy2019vol1,candy2019vol2} that rarely find themselves in conversation. 
Encouraging discussion among the diverse researchers keen on testing behavioral hypotheses about future technologies could lead to the fertile recombination of disciplinary expertise, significantly increasing the methodological toolkit of this new community.

Still, most of these solutions will be focused on the first threat to prospective validity, that is, the challenge of eliciting responses from participants that accurately comport to how people would react in an anticipated future. 
They do not address the second and third challenges of the difficulty in accurately anticipating a technology and the social context in which it will be deployed. 
Here researchers need to balance the benefits that their research could bring with the uncertainty inherent in a speculative, prospective science. 
Nevertheless, certain topics will be laden with more uncertainty than others. 
Threats to prospective validity should thus be minimized by recognizing the features that can make topics less speculative and more amenable to study.  
In the next section, we consider some first guidelines for choosing such topics.

\section*{Topics for science fiction science}

In the science fiction novel \emph{Too Like the Lightning,} author Ada Palmer \cite{palmer2016too} describes a future where the hyper-mobility afforded by free, flying, supersonic self-driving cars led to the collapse of nation states, replaced by global communities of like-minded individuals. 
While this technology makes for a fantastic story, it would arguably be inadequate as the focus of a sci-fi-sci experiment. 
To start with, the development of such technology is very far from our current capabilities. 
This makes it difficult to imagine its plausible specifications, and even harder to envision the societal landscape by the time it might be invented. 
In addition, geopolitical effects like the collapse of nation states unfold over a long period of time (years or even generations). 
Such effects cannot be properly measured within the brief duration of an experiment. 
Finally, the target technology is so disruptive that it transforms everything.
This makes it difficult for participants to simulate the behavior of future people, if they not only have to picture a new technology, but also deal with its all-encompassing ramifications. 

\begin{figure*}
\includegraphics[width = \textwidth]{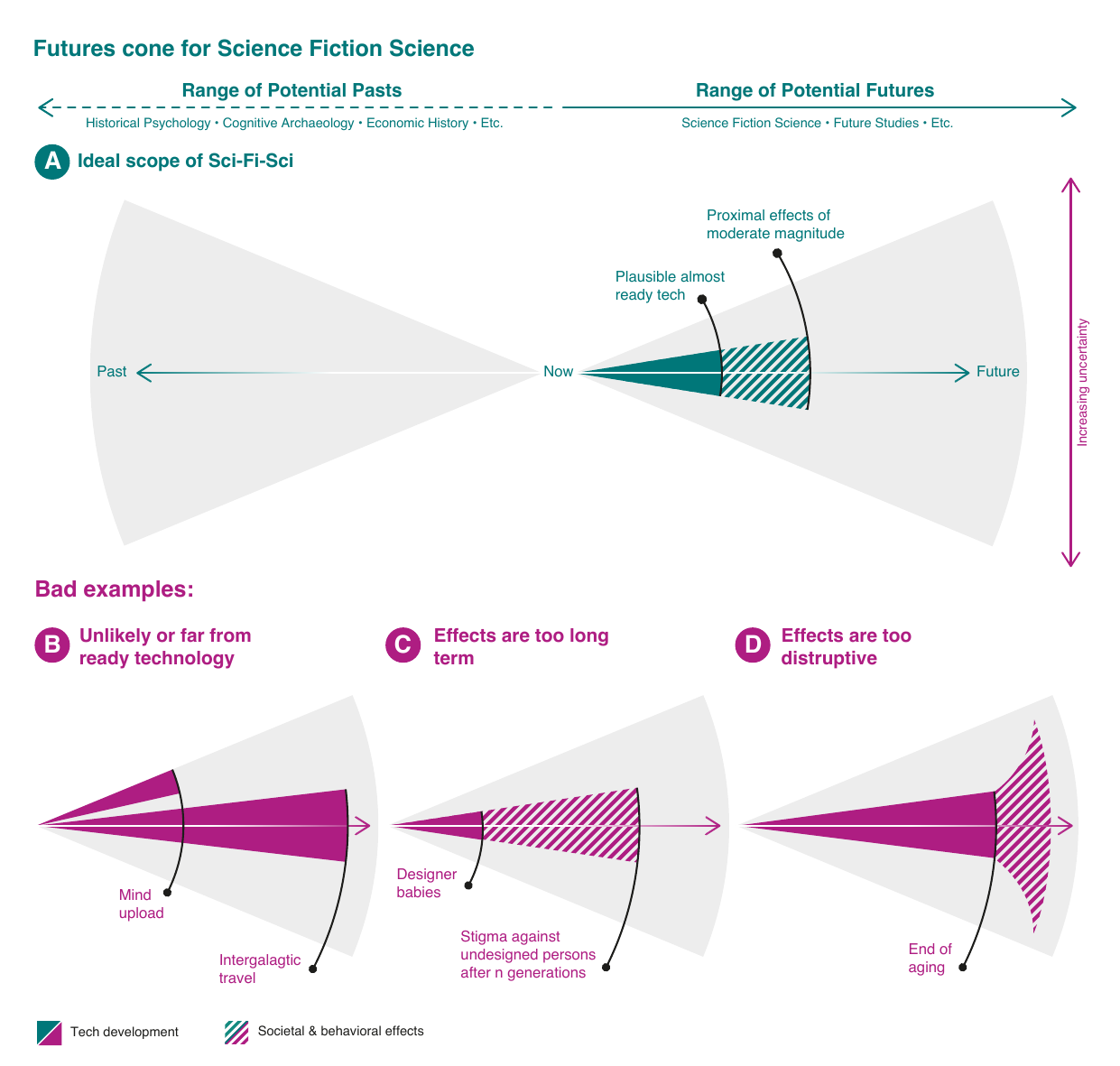}
\caption{\textit{A modified `Futures' cone \cite{gall2022visualise} for Science Fiction Science. \textbf{(A)} The ideal domain for sci-fi-sci is to study proximal, moderate magnitude effects of plausible, almost ready technology. \textbf{(B--D)} show different deviations from this ideal: Topics less amenable to sci-fi-sci include unlikely or far from ready technologies \textbf{(B)}; societal and behavioral effects that only materialize too far into the future after the introduction of the technology \textbf{(C)}; and technologies whose effects that too disruptive that they would change so many aspects of society all at once \textbf{(D)}.}}\label{fig:cone}
\end{figure*}

The above example illustrates how uncertainty around the future can become simply too large to conduct a valid experiment. This concept is often captured visually via a so-called \emph{futures cone} \cite{gall2022visualise}, which we have adapted for our purposes (Figure \ref{fig:cone}).
The cone of uncertainty represents the idea that, as we project further into the future (or the past), the range of possible outcomes (or histories) broadens due to the increasing number of uncertain or unpredictable variables. This is true in fields that study past human behavior (e.g. cognitive archaeology \cite{coolidge2015cognitive}, historical psychology \cite{gall2022visualise} and economic history \cite{north1978structure}) as well as future behavior (e.g. sci-fi-sci and Future Studies). We can adapt the futures cone to explore what makes an ideal topic for sci-fi-sci.

To start with, the further away a technology is in the future, the greater the cone of uncertainty. 
As a result, it is good practice to stick to the near future when designing a sci-fi-sci experiment.
What constitutes the `near future' is up for debate, of course. 
But one place to start is by studying the near-term behavioral effects of applications which seem to be almost within reach. 
For example, recent advances in Large Language Models may enable widespread use of automated lie detection in interpersonal communication. Behavioral experiments can explore how such capabilities may disrupt existing social dynamics---e.g., making people less inhibited in accusing others of lying \cite{von2024lie}.
When it comes to technologies that do not yet exist, one heuristic is to consider the Technological Readiness Level (TRL) of the target technology \cite{heder2017nasa}.
The TRL scale, developed by NASA and adopted by many other agencies, assesses the maturity of a technology. 
It ranges from level 1, where only basic scientific principles are observed and potential applications are not well formulated, to level 9, where the technology is validated in its intended operational environment. 
While sci-fi-sci studies may not need to wait until a technology reaches level 9, it seems reasonable to focus on technologies that are at least 4 (small scale prototype) on this scale.
The further along a technology is on the TRL scale, the more likely its sci-fi-sci version will resemble its eventual real version.

In addition, when running a sci-fi-sci experiment, we must consider not only when a technology will appear but also the time scale of the effects we want to investigate. 
The longer it takes for these effects to unfold, the larger the cone of uncertainty around them, and the greater the threat to validity. 
Therefore, it is probably good practice to study the proximal effects of the technology rather than those that unfold over a time scale beyond that of a behavioral experiment.
For example, we could study the social acceptability of future parents selecting or deselecting various traits when choosing an embryo for implantation. 
However, it would be much more difficult to study how their choices might affect the social stigma attached to the deselected traits over time. 

Finally, there are some technological breakthroughs (e.g., anti-aging treatments that could push human longevity into centuries) that would be so disruptive that their appearance may change everything, exponentially increasing the cone of uncertainty around them.
While science fiction writers may naturally gravitate toward these game-changing technologies, because of their narrative potential, sci-fi-sci experimenters may prefer to focus on technologies that bring changes of moderate magnitude. Although there is no analogue to the TRL for quantifying the predicted impact of a future technology, Coccia’s \textit{Scale of Innovative Intensity} \cite{coccia2005measuring} categorizes \textit{historical} technologies into three tiers of impact: low, medium, and high. The high-impact category is further divided into `very strong' (e.g., the Internet) and `revolutionary' (e.g., steam power, electricity) technologies with such profound effects that they reshape nearly every sector of the economy and touch nearly every individual on the planet. Again, as a rough heuristic, sci-fi-sci may be best targeted at emerging technologies that would fit in the low- and medium-impact tiers of the Coccia scale. Recent examples of technological innovations in this range include personalized algorithms for media content, direct-to-consumer genetic testing, SMS texting, credit cards and other forms of digital cashless payment, 24-hour news channels, and in-vitro fertilization. Each of these was an incremental technological advance that could have been predicted in the years preceding its emergence, but each also had pronounced and often unintended societal consequences.   

This is not to say that sci-fi-sci should fully avoid sweepingly transformative technologies like AI, which will almost certainly have `very strong' and possibly `revolutionary' impact. Sci-fi-sci can and should engage with AI, yet it is challenging to design experiments around the technology's most ambitious, civilization-changing scenarios. Instead, sci-fi-sci researchers are better off breaking down AI into more manageable components and narrower applications, such as self-driving technology or autonomous medical triage---just as researchers in the near past would have found it more manageable to design experiments around the impact of a transformative but evolutionary technology such as social media rather than a paradigm shifting technology like the Internet (or electricity). Such scenarios remain close enough to current conditions that researchers can realistically simulate them, vary key parameters, and measure meaningful responses without collapsing under the weight of unbounded speculation. In this sense, we do not propose sidestepping AI altogether, but rather targeting more delimited forms of AI-driven technologies whose behavioral impacts can be credibly tested in a laboratory context.

While it is difficult to define formal guidelines for topic selection (in the broad sense of choosing a technology of interest, and defining the exact scenarios for investigation), it is an area where synergy between sci-fi-sci and futures studies will be especially important. Indeed, futures studies researchers have already established a trove of ready topics through methods such as backcasting, whereby futurologists trace back the technological path (and societal response) that would lead to a desirable outcome. Their topics have been qualitatively derived, and now stand ready for quantitative testing via the sci-fi-sci method.
Moreover, in order to generate realistic and comprehensive scenarios grounded in domain-specific knowledge, sci-fi-sci researchers will find great value in collaborating with governmental and non-governmental organizations that engage in scenario planning; with experts of methods such as prospective hindsight and pre-mortem studies; and with practitioners who help ensure that experimental designs capture critical variables and potential harms that might otherwise be overlooked. 
By leveraging the detailed narratives and insights from futures studies, sci-fi-sci can create more meaningful and effective experiments, enhancing its ability to anticipate and study the impacts of speculative technologies.

\section*{Final Remarks}

Science fiction writers give us a glimpse of possible futures, by telling the stories of how humans might be changed by new technologies. 
Sci-fi-sci applies the scientific method to the science fiction project, by immersing research participants into controlled variations of the future, and collecting quantitative data on their attitudes and behavioral responses. 
The seeds of sci-fi-sci were planted before this article:  researchers from various fields are already providing immersive simulations of the future, or asking research participants to express preferences and opinions about the technologies of the future. 
Accordingly, we do not claim to have invented a new field, but we hope that researchers who are already studying the behavior of future humans, in disparate fields, will have an easier time finding each other with a flag to rally around.

We also hope that new research communities will discover the sci-fi-sci project, and bring their expertise to this multidisciplinary enterprise.
Studying the behavior of future humans interacting with future technology in a future social world raises unusual challenges for behavioral scientists, which call for unconventional methods. Institutional mechanisms that can help overcome these challenges include: higher risk tolerance among research funding organizations to create programs explicitly designed for sci-fi-sci projects despite the substantial uncertainties involved; community building meetings to facilitate the sharing of experiences and best practices; and coordinated efforts to establish rigorous standards for assessing sci-fi-sci experimental methods and findings. In particular, sci-fi-scientists will need to devise best practices to communicate their quantitative findings, so as to avoid conveying a false sense of precision to stakeholders who may overfixate on numerical estimates that come with large uncertainty.

We hope that our review of these methods, their rationales and their limitations, will encourage sci-fi scientists to make bold methodological choices, and a convincing case for these bold choices.
The future of science fiction science will depend on the quality of its methods.

\section*{Contribution Statement}
The conceptualisation and writing of this article was a joint effort of all three authors.

\section*{Competing Interests}
The authors declare no competing interests.

\section*{Acknowledgments}
JFB acknowledges support from grant ANR-19-PI3A-0004, grant ANR-17-EURE-0010, Grant ANR-22-CE26-0014-01, and the research foundation TSE-Partnership. AS acknowledges a Canada 150 Research Chair from the Social Science Research Council of Canada.


\end{document}